\documentclass[pre,twocolumn,aps,amsmath,amssymb,superscriptaddress]{revtex4}

\usepackage{graphicx,subfigure}
\usepackage{amssymb}
\usepackage{amsmath}
\usepackage[british]{babel}
\usepackage{color}

\begin{document}

\title{Critical adsorption of polyelectrolytes onto charged Janus nanospheres}

\author{Sidney J. de Carvalho}
\affiliation{Department of Physics, S\~ao Paulo State University, S\~ao Paulo,
Brazil}
\author{Ralf Metzler}
\affiliation{Institute for Physics and Astronomy, University of Potsdam,
Potsdam-Golm, Germany}
\affiliation{Department of Physics, Tampere University of Technology, 33101
Tampere, Finland}
\author{Andrey G. Cherstvy}
\affiliation{Institute for Physics and Astronomy, University of Potsdam,
Potsdam-Golm, Germany}
\email{Corresponding Author: a.cherstvy@gmail.com}

\begin{abstract}

Based on extensive Monte Carlo simulations and analytical considerations we study the electrostatically driven adsorption of flexible polyelectrolyte chains onto charged Janus nanospheres. These net-neutral colloids are composed of two equally but oppositely charged hemispheres. The critical binding conditions for polyelectrolyte chains are analysed as function of the radius of the Janus particle and its surface charge density, as well as the salt concentration in the ambient solution. Specifically for the adsorption of finite-length
polyelectrolyte chains onto Janus nanoparticles, we demonstrate that the critical adsorption conditions drastically differ when the size of the Janus particle or the screening length of the electrolyte are varied. We compare the scaling laws obtained for the adsorption-desorption threshold to the known results for uniformly charged spherical particles, observing significant disparities. We also contrast the changes to the polyelectrolyte chain conformations and the binding energy distributions close to the adsorption-desorption transition for Janus nanoparticles to those for simple spherical particles. Finally, we discuss experimentally relevant physico-chemical systems for which our simulations results may become important. In particular, we observe similar trends with polyelectrolyte complexation with oppositely but heterogeneously charged proteins.\\

Abbreviations: ES, electrostatic; PE, polyelectrolyte; JNS, Janus nanosphere;
WKB, Wentzel-Kramers-Brillouin.
\end{abstract}

\maketitle

\section{Introduction}

\begin{figure}
\includegraphics[width=8cm]{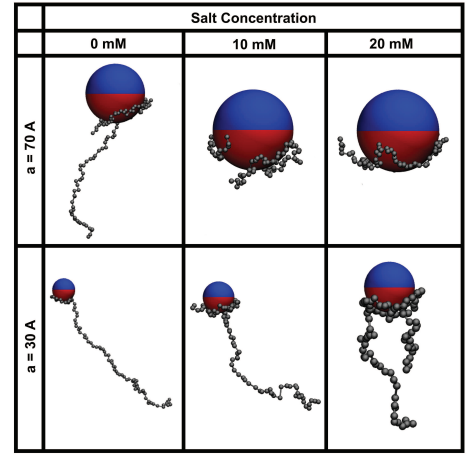}
\caption{Typical conformations of PE chains in front of JNS of radii $a=70$ \AA
 \ (top) and  $a=30$ \AA\  (bottom), as obtained from our computer simulations
(particles are shown not to scale). The salt concentration from left to right
is 0, 10, and 20 mM and the particle surface charge density is $\sigma a^2
\approx 7.5$.}
\label{fig-config}
\end{figure}

Janus grains \cite{Degennes:1992} or particles of spherical, cylindrical, or
disc-like architecture are nanoparticles whose surface is divided into two
or more areas with distinct physico-chemical properties. For instance, the
behaviour of Janus particles with one apolar hydrophobic half and a polar
or charged half \cite{Walther:2008} share common features with surfactant
assembly in solutions \cite{Granick:2008}. Charged Janus nanospheres (JNSs),
our main focus in this study, exhibit two oppositely charged hemispheres. We
study in detail the critical adsorption conditions for a polyelectrolyte chain, see Fig.~\ref{fig-config}. From extensive Monte Carlo simulations as well as analytical calculations we
demonstrate that the adsorption behaviour to JNS significantly differs from
that to homogeneously charged spheres. The polyelectrolyte adsorption is
investigated with respect to changes in the size of JNSs and salinity of the electrolyte.

Janus particles have become widely used model systems. Thus, orientation-sensitive
interactions between JNSs \cite{Roij:2012} are extensively used in directed
supra-molecular hierarchical assembly \cite{Granick:2009,Granick:2010}. Janus
particles are promising candidates for designing novel materials, responsive
sensor nano-devices, biomedical and pharmaceutical applications, anti-reflection
coatings, stabilising liquid-liquid interfaces, as well as nanoscale chemical
locomotion \cite{Walther:2008,Glaser:2006,Chen:2012,Rachel:2011}. ``Hairy'' Janus
particles decorated via grafting \cite{Szleifer:2012} of various polymers
\cite{Chen:2012} reveal rich stimuli-responsive behaviour \cite{Berger:2008} and
can serve as coatings of immiscible liquid-liquid interfaces \cite{Muller:2008,
Krausch:2006}.

The weak and strong adsorption of polyelectrolytes (PEs) onto oppositely charged
interfaces has a number of important technological and biological applications
\cite{Winkler:2013}. These include, inter alia, paper production, various surface
coating techniques, colloid stabilisation, adsorption of nucleic acids on the
interior of viral shells, DNA wrapping in nucleosomes, pharmacological
applications, etc. Of particular interest is the regime of weak adsorption,
referring to weakly charged PEs below the Manning's counterion condensation limit
\cite{Manning:1978,Manning:2007}, i.e., charge-charge distances along the PE
contour are shorter than the Bjerrum length $l_B$. A fundamental feature of weak
PE adsorption is the transition between the entropically driven free state of the
PE chain in solution and the bound state close to the interface. The latter is
favoured by electrostatic (ES) PE-surface attractions, which are typically screened
exponentially by the surrounding electrolyte in a Debye-H{\"u}ckel fashion. This
transition defines the condition for the \textit{critical adsorption\/} of PE
chains, at which the entropic penalty of PE confinement near the interface is
exactly balanced by the energetic attraction to the surface. At larger electrolyte
salinities the PE-surface attractions are better screened and thus larger surface
charge densities are necessary to trigger the adsorption. The study of this transition
for dipolar JNSs is the main target of the present study, as depicted in
Fig.~\ref{fig-config}.

Theoretically, the adsorption of PE chains onto uniformly charged surfaces was
studied to great detail, see, e.g., Refs.~\cite{Netz:2003,Dobrynin:2005,
Dobrynin:2008,Winkler:2013,Shafir:2004,Netz:2005,Netz:1999,Schiessel:2000,
Sens:2000,Schiessel:2003,Messina:2009,deVries:2006,Rudi:2006,Borkovets:2014,
Netz:2014,Metzler:2014}. The regime of weak PE-surface adsorption for which the
entropy of the PE chain is still important, was extensively studied by computer
simulations \cite{Kong:1998,Linse:2002,Dobrynin:2011,Nandy:2011,Cao:2013,
Nunes:2013,Molina:2013,Molina:2014,Man:2010,Wang:2011,Eli:2012,Krefeld:2013,
deCarvalho:2013}. The characteristics of the PE-surface adsorption including the
conditions for the critical adsorption of long PE chains with a constant linear
charge density onto planar \cite{Wiegel:1977,Muthukumar:1987} and curved
\cite{vanGoeler:1994,Baumgaertner:1991,Netz:1999-2,Winkler:2006,Cherstvy:2006,
Winkler:2007,Cherstvy:2011,Muthu:2011,Winkler:2013,Cherstvy:2012,
Cherstvy:2012dielectric,deCarvalho:2010,Forsman:2012} interfaces were also
derived. Apart from the more conventional strong PEs, the adsorption of weak
charge-regulated PEs and poly-ampholyte chains onto spherical particles was
investigated by computer simulations \cite{StollWeak:2011,Stoll:2011,
StollPolyamph:2005} as well as experiment \cite{Chen:2011,Kayitmazer:2013}.

The paper is organised as follows. In Sec.~\ref{sec-critical-intro} we set the
scene by briefly reviewing the main results for the critical PE adsorption onto
planar and curved surfaces. Sec.~\ref{Sec_MC} elucidates the PE adsorption onto
JNSs and discusses the core results of the current paper obtained from extensive
Metropolis Monte-Carlo computer simulations. We vary the solution salinity, the
particle radius, and the length of polymer chains to unveil their impact on the
location of the adsorption-desorption boundary. In Sec.~\ref{Sec_Disc} we discuss
our results, outline their possible applications, and draw our conclusions.
App.~\ref{sec-app-es} details the exact solution of the ES potential around a
JNS within the linear Poisson-Boltzmann model. In App.~\ref{Sec_WKB} the equations
for the 2D WKB theory of PE adsorption are derived.

\section{Critical polyelectrolyte adsorption}
\label{sec-critical-intro}

We here briefly review the results for the critical PE adsorption onto planar and
curved interfaces necessary for our discussion regarding PE-JNS adsorption below. The PE adsorption
occurs when the gain of the PE-surface ES attraction becomes larger than the free
energy loss due to the decrease in the conformational entropy of the adsorbed
PE chain. Critical adsorption conditions relate the surface charge density
$\sigma$, the PE linear charge density $\rho$, the ambient temperature $T$, the
persistence length $b/2$ of the PE chain, the radius of curvature $a$ of the
surface, and the electrolyte concentration. These quantities are connected by
the single universal adsorption parameter, denoted below as $\delta$. The precise form of this coupling is
\cite{Cherstvy:2011}
\begin{equation}
\delta_c=24\pi\frac{|\sigma_c\rho|}{\epsilon k_BTb}a^3,
\label{def_delta}
\end{equation}
where the subscript $c$ denotes the value at the critical adsorption point.

The exact value of $\delta_c$ decides whether for given physical parameters a PE
chain adsorbs onto the surface or stays unbound. In the particular case of a
\emph{planar\/} interface the exact analytical result for long flexible PE chains
shows that the critical surface charge density $\sigma_c$ scales with the third
power of the reciprocal Debye screening length $\kappa=1/\lambda_D$
\cite{Wiegel:1977}, so that the approximate result is
\begin{equation}
\delta_c^{pl}(\kappa)\approx1.45\left(\kappa a\right)^3.
\label{delta_plane}
\end{equation}
This result is getting modified substantially for more complex surface geometries and its physical properties. For instance, larger plane surface charge densities are necessary to induce PE adsorption when the surface is low-dielectric \cite{Cherstvy:2012dielectric}.
Due to PE-PE image-force repulsion \cite{Sens:2000} in the course of adsorption,
substantially larger critical strengths $\delta_c$ are required to trigger
polymer adsorption. In addition, in the low-salt limit the cubic scaling $\delta_c\sim\sigma_c\sim
\kappa^3$ given by expression (\ref{delta_plane}) is modified to the quadratic
scaling $\delta_c\sim\kappa^2$ \cite{Cherstvy:2012dielectric}.

For \emph{curved\/} convex surfaces, higher values of $\sigma$ are necessary to
activate the adsorption, as a larger entropic penalty needs to be paid for the
polymer confinement near the surface, meaning that the value of $\delta_c$
becomes larger than $\delta_c^{pl}$. Moreover, the scaling $\delta_c(\kappa)$
changes with the surface curvature and the salinity of the electrolyte. These
two quantities can be combined into a dimensionless parameter $\kappa a$
controlling the scaling behaviour $\delta_c(\kappa a)$ for long chains.
Specifically, in the limit of low salt or large surface curvature, when $\kappa a
\ll 1$, it was shown recently that for a \emph{spherical particle\/} the linear
dependence
\begin{equation}
\delta_c^{sp}(\kappa)\approx0.90\left(\kappa a\right)^1.
\label{delta_sphere}
\end{equation}                          
emerges \cite{Cherstvy:2011}. For otherwise given parameters, this relation
determines the minimal sphere radius $a_c$ required for PE adsorption as well as
the critical adsorption temperature $T_c$. For flexible PEs in front of a
\emph{cylindrical interface\/} the scaling relation becomes
\begin{equation}
\delta_c^{cyl}(\kappa)\approx1.07\left(\kappa a\right)^2
\label{delta_cyl}
\end{equation}
in the limit $\kappa a \ll 1$ \cite{Cherstvy:2011}. This change in the exponent
for spherical versus cylindrical particles is corroborated by experimental data
\cite{Feng:2001HAHAHA} (see also Fig.~11 and the discussion in
Ref.~\cite{Cherstvy:2006}).

To derive the scaling relations (\ref{delta_sphere}) and (\ref{delta_cyl}), the
Wentzel-Kramers-Brillouin (WKB) approximation scheme known from quantum mechanics
was employed \cite{Cherstvy:2011}. It accurately reproduces the known results,
including the pre-factors, for $\delta_c(\kappa a)$ in planar \cite{Wiegel:1977} and
spherical \cite{Winkler:2006} geometries. In the opposite limit of $\kappa a\gg1$
the WKB theory predicts that for convex surfaces $\sigma_c$ approaches the value
for the uniformly charged plane from above, and the scaling (\ref{delta_plane})
is valid \cite{Cherstvy:2011}. The reasons for this are a) the sphere and cylinder
ES potentials are lower than the one for the planar surface with identical $\sigma$ and b)
the polymer conformations are perturbed more strongly on adsorption onto convex
surface-limited interfaces such as those of spherical and cylindrical particles. This reduction of the number of polymer translational degrees of freedom requires larger $\sigma_c$, particularly at low salinities and small $\kappa a$ values. Concurrently,
in the low-salt limit a systematic change in the scaling of the critical adsorption
parameter $\delta_c$ was predicted \cite{Cherstvy:2011}, in qualitative agreement
with experimental observations \cite{Kayitmazer:2013}, see Sec.~\ref{Sec_Disc}.
Extensive comparison of theoretical predictions for the critical PE adsorption
onto spherical versus cylindrical surfaces to a number of experimental data sets was presented in Refs.~\cite{Cherstvy:2006,Winkler:2013}.  

These theoretical predictions for critical PE adsorption were corroborated indirectly by experiments, in that complex formation and coacervation of PEs with oppositely charged particles revealed no
adsorption and agglomeration for $\kappa$ above the critical value, when strong
screening effects significantly reduce the PE-particle ES attraction
\cite{Chen:2011,Kayitmazer:2013,Dubin:1992,Feng:2001HAHAHA,Dubin:2010,
Cooper:2005,Cooper:2006,Kizilay:2011,Dubin:2001powers}. The critical surface
charge density $\sigma_c\sim\kappa^\nu$ for spherical and cylindrical particles
was demonstrated to increase with $\kappa$ as follows \cite{Dubin:2001powers}
\begin{equation}\sigma_c^{sp}(\kappa)\sim\kappa^{1\dots 1.8} \mbox{ and }
\sigma_c^{cyl}(\kappa)\sim\kappa^{1.8\dots 2.5}.
\label{eq-dubin-scaling}
\end{equation}
The scaling exponents $\nu$, being sensitive functions of PE stiffness and linear
charge density, clearly differ from the cubic dependence (\ref{delta_plane}) due
to curvature effects of adsorbing interfaces as well as the finite available
adsorption surface and salt-dependence of PE persistence length $l_p$.

These critical adsorption results for uniformly charged interfaces have to be modified in many
realistic situations when quenched or annealed charge patterns exist on the
adsorbing surfaces. This may be due to the discrete nature of surface
charges or some inherent non-uniform or rough structure of interfaces. For
\textit{planar patterned\/} surfaces, several theoretical \cite{McNamara:2002,
Hoda:2008} and in silico \cite{Muthukumar:1995,Ellis:2000,deVries:2008,Dias:2005}
studies are available which focused on the effects of charge inhomogeneities on
PE adsorption. For instance, for a circular charged patch of radius $R$ the critical PE
adsorption condition was derived in a variational model \cite{Muthukumar:1995}.
As expected, a patch of a finite size demands higher $\sigma$ values to enable
adsorption \cite{Muthukumar:1995,Variational}. Effects of the patch size and
charge distribution for polyion-protein complexation were analysed by computer
simulations for complexes close to the isoelectric point \cite{deVries:2004},
see also Ref.~\cite{Carlsson:2001}. Computer simulations for planar patterned
charged interfaces were also reported \cite{Hoda:2008}. 

For \textit{curved patterned\/} interfaces, however, the understanding of the
critical PE adsorption remains elusive. It is the main purpose of the present
study to advance the knowledge on this process. We focus on PE adsorption onto
JNS depicted in Fig.~\ref{fig-config} which is the ultimate example of a
net-neutral patchy colloid \cite{Walther:2008}. It consists of two equally but
oppositely charged hemispheres so that the area of charged patches and surface
curvature are intimately coupled. We show that this surface charge distribution
effects a non-trivial interplay of the patch-size-mediated adsorption-desorption
and the particle size, adds interesting new insight to the theory of PE adsorption
onto curved interfaces, and represents a first step towards a more general
approach to less orderly charge distributions.

In what follows we use Metropolis Monte-Carlo computer simulations to quantify the
conditions of the critical PE adsorption as a function of the surface charge
density and curvature of JNSs, the salinity of the solution, and the length
of PE chains. The simulations are based on an algorithm, which was previously
successfully implemented by one of the author to study the PE adsorption onto spherical and cylindrical
uniformly-charged interfaces \cite{deCarvalho:2010,deCarvalho:2013}.

\section{Simulations of PE-JNS adsorption}
\label{Sec_MC}

\subsection{Specification of the model and the employed parameters}

A single JNS of radius $a$ is fixed at the centre of a spherical simulations box
containing a single PE chain. The latter is modelled as a bead-spring polymer
consisting of $N$ single-charged spherical monomers connected by harmonic elastic
potentials
\begin{equation}
E_{el}(\delta r_{i,j})=\frac{1}{2}kr^2_{i,j},
\end{equation}
and interact via the screened Coulomb ES potential with a hard-core restriction,
\begin{equation}
E_{ES}(r_{i,j})=e_0^2\frac{\exp(-\kappa r_{i,j})}{\epsilon r_{i,j}}.
\end{equation}
Here $r_{i,j}$ is the distance between monomers $i$ and $j$ with $1>(i,j)>N$ and
$i\neq j$, and we set $e_0=1$ below. This inter-monomer ES repulsion yields a
\textit{finite\/} persistence length of the polymer that will be a crucial
ingredient for the critical adsorption conditions derived in Sec.~\ref{sec-crit}.
We thus explicitly account for the intra-chain charge-charge interactions for
finite-length chains, a situation which would require a formidable theoretical
modelling. The chain parameters are the same as used in Ref.~\cite{deCarvalho:2013}
to study the PE-cylinder adsorption. The screened monomer-monomer interactions
imply a finite PE ES persistence length, $l_p^{el}$, in contrast to the ideal
Gaussian chains often employed in theoretical approaches  \cite{Cherstvy:2011}. 

The ES interactions of the $i$th PE monomer with the JNS is governed by the
attractive dipolar-like potential $\Psi_J$ given by Eq.~(\ref{exact_potential}) in the
Appendix, see also Fig.~\ref{dipole} for the ES potential distribution. PE-JNS ES binding energy then is
\begin{equation}
E_{B}(r_i,\theta_i)=e_{0}\Psi_J(r_i,\theta_i).
\label{eq-ebinding}
\end{equation}
Simulating the PE adsorption onto a uniformly charged sphere,
Eq.~(\ref{eq-pot-sphere}) for $\Psi_{sp}(r)$ is used. The aqueous solution has the standard
dielectric constant $\epsilon=78.7$ and temperature  $T=298.15$K. The
monomer radius is $r=2$ \AA, and each monomer carries the elementary charge
$e_0$. The system is confined inside a large electrically-neutral simulations
cell to eliminate end- and finite-size effects. 

\begin{figure}
\includegraphics[width=7cm]{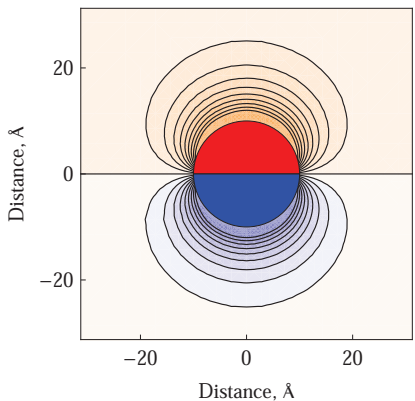}
\caption{Two-dimensional contour-plot of the ES potential $\Psi_J(r,\theta)$
around the JNS as given by Eq.~(\ref{exact_potential}). Blue represents the
positive and red the negative values of the potential. Parameters: $a=10$ \AA,
$1/\kappa=10$ \AA, and $S_0=300$ \AA$^2$ .}
\label{dipole}
\end{figure}

To sample the polymer configurations, translational movements of individual
monomers and of the whole chain are combined with the Pivot rotation, see the
detailed description in Ref.~\cite{deCarvalho:2013}. A Metropolis Monte-Carlo
approach in the canonical ensemble is implemented to explore the configurations
space of the polymer-particle system. The standard Weighted Histogram
Analysis Method (WHAM) is used to compute the system free energy. The polymer profiles are computed from the distribution functions of its monomers and their centre-of-mass $\mathbf{R}_{CM}(t)$ generated, as a function of the distance from the macro-ion surface of JNSs, $p(\mathbf{ R}_{CM}-\mathbf{R}_{m})$.

\subsection{Adsorption criterion}

The criterion for the PE-particle adsorption is based on the following
considerations. As we observed for the interaction of a PE chain with a
uniformly charged sphere, at higher ionic strengths the particle-PE attraction
cannot overcome the loss of conformational and translational polymer entropy,
and thus no  adsorption takes place \cite{deCarvalho:2010}. With a decreasing salinity, a \emph{discontinuous},
first-order transition between the adsorbed and desorbed states of the PE chain
occurs. The adsorbed state is characterised by binding energies of, typically,
several units of the thermal energy $k_BT$, whereas the desorbed state has zero
energy. This ``gap'' allows us to count the fraction of polymer configurations generated
from simulations in each state, without defining an (arbitrary) energy threshold or a
shell around the adsorbing particle to distinguish adsorbed versus desorbed PE
monomers. Following this observation, we count the polymer as adsorbed if more than 50\% of
the configurations are in the adsorbed state during the simulation time. Typically
of the order of $10^6$ to $10^7$ polymer configurations are used for averaging.
As we show below, for PE-JNS adsorption there exists indeed a similar critical
energetic transition between the adsorbed and desorbed states.

\begin{figure}
\includegraphics[width=7cm]{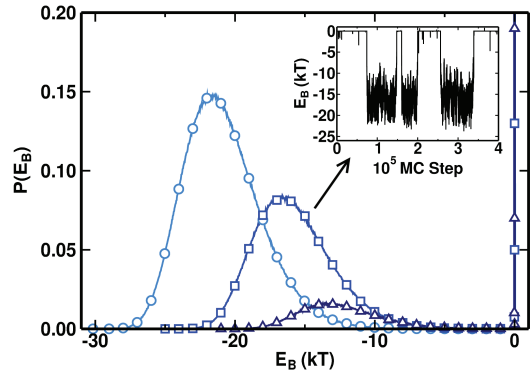}
\caption{Distribution $p(E_B)$ of the ES binding energy obtained
from simulations of PE-JNS adsorption at varying salinities. The inset
shows the temporal evolution of the binding energy $E_B(t)$, tracing transitions
between the adsorbed and desorbed states of the PE chain, in units of the thermal
energy $k_BT$. The profiles are normalised to unity when the zero binding energy
contributions stemming from the desorbed states (shown as vertical lines at $E_B=0$) are considered as well.
Parameters: the JNS surface charge density is $\sigma
=25$mC/m$^2$ \textcolor{black}{(bluish colours for the curves)},
salt concentration is 20 (circles), 30 (squares)  and 40 mM (triangles),
the JNS radius is $a=$70 \AA, and the number of PE monomers is $N=20$.}

\label{fig-fe}
\end{figure}

\begin{figure}
\includegraphics[width=7cm]{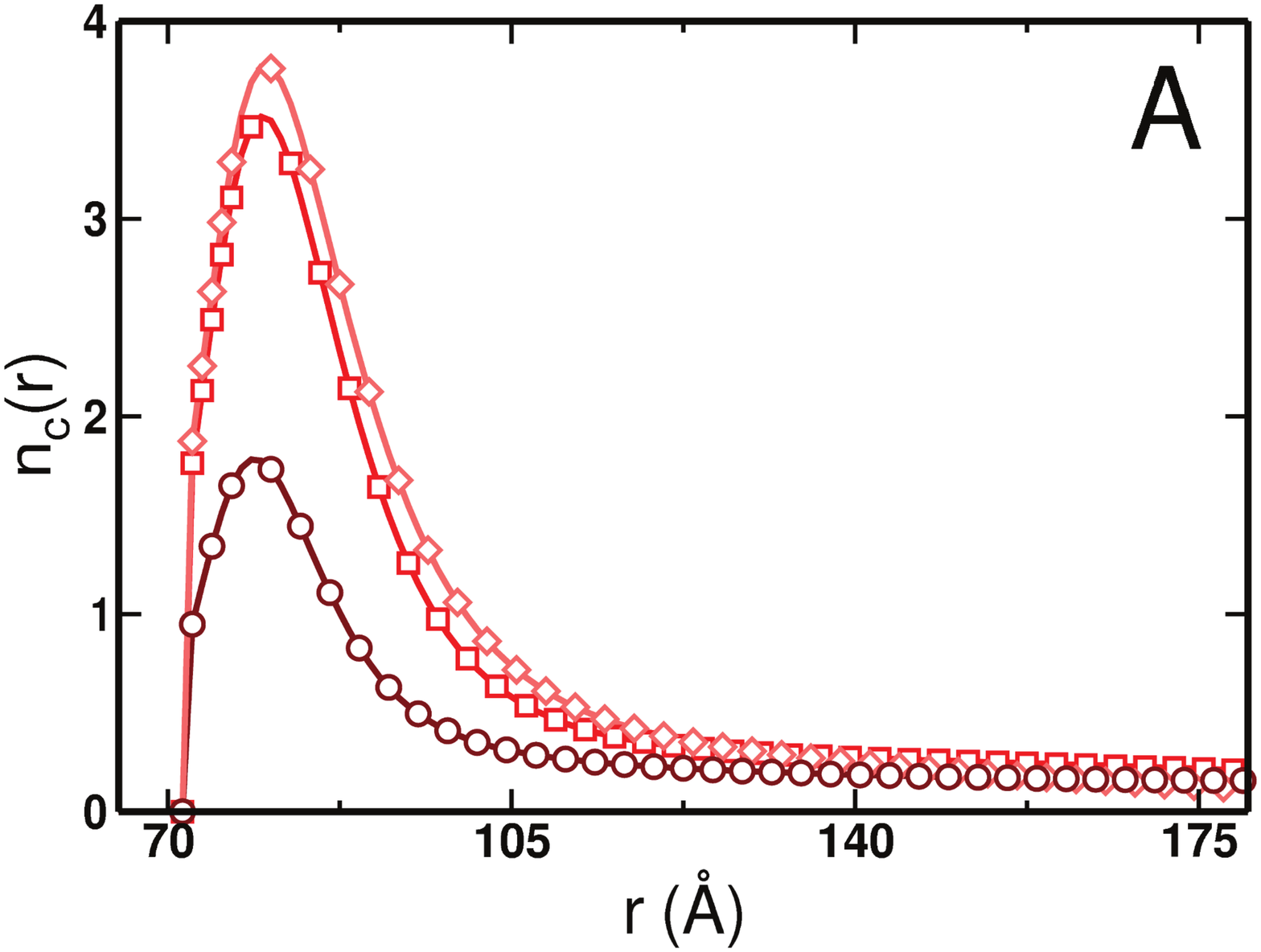}
\includegraphics[width=7cm]{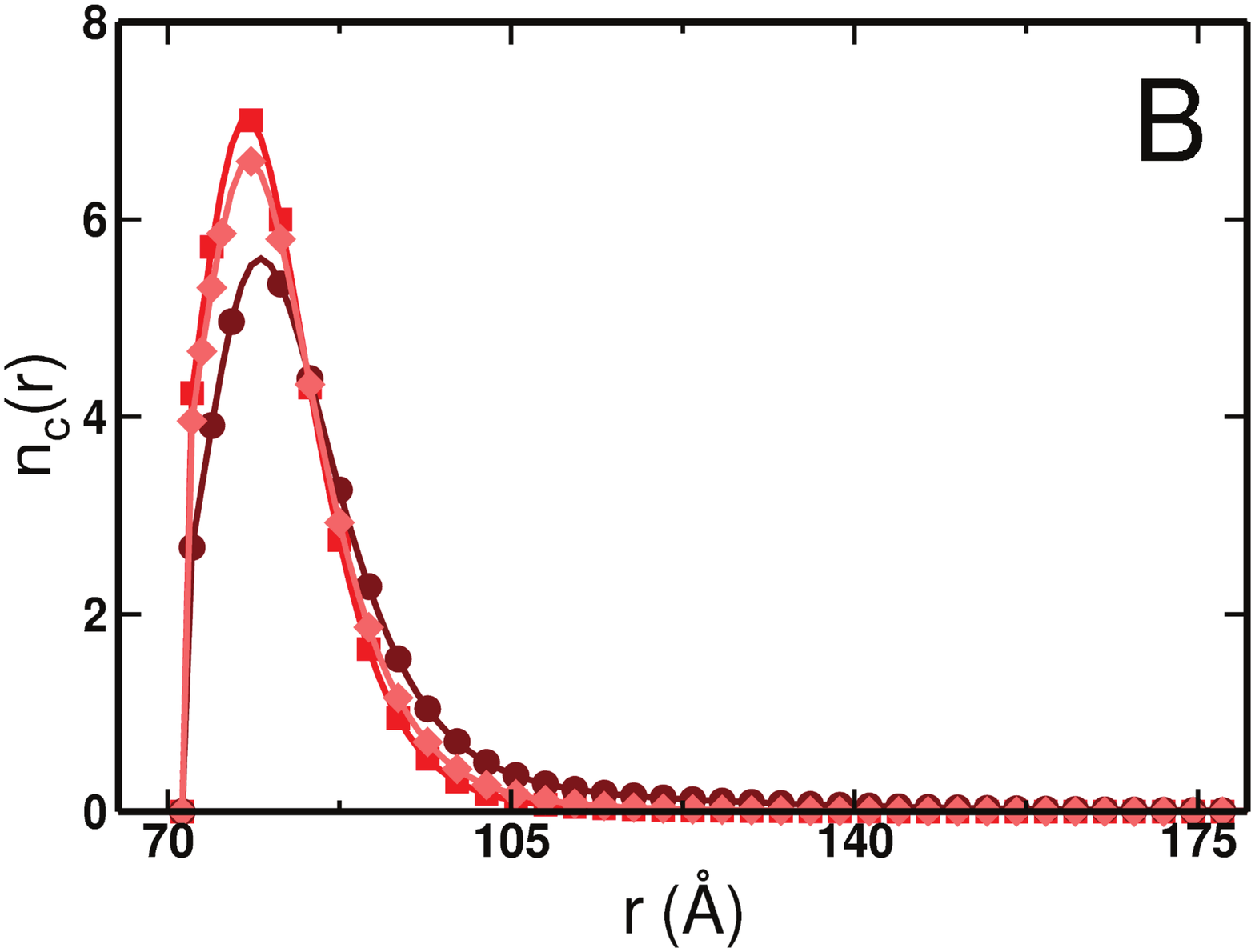}
\caption{Number of adsorbed PE monomers $n_c(r)$ versus distance from the
adsorbing JNS (panel A, open symbols) and from the uniformly charged
sphere (panel B, filled symbols). Parameters: $\sigma=12.5$ mC/m$^2$
\textcolor{black}{(reddish colours for the curves)}, salt
concentration is 0 (circles), 6 (squares) and 14 mM (rhombi), $a=70$ \AA,
and $N=100$.}
\label{fig-profile}
\end{figure}

\section{Main Results}

\subsection{Binding energies and distributions of PE monomers}

Typical conformations of the PE chain in the proximity of JNSs of different
sizes and at varying salinity of the solution are presented in
Fig.~\ref{fig-config}. The differences for varying conditions are striking,
ranging from a completely adsorbed chain to a configuration, in which the chain
is marginally attached to the JNS and points almost completely straight away from
it. We observe that at higher salinities and larger Janus spheres a more pronounced
PE adsorption takes place. This trend with the solution salinity is opposite to that for PE-sphere adsorption, where the adsorption gets amplified in low-salt conditions. For PE-JNS adsorption at low salt, the ES PE attraction by the oppositely charged hemisphere is accompanied and compensated by ES repulsion by the similarly charged Janus hemisphere, reducing the number of polymer monomers adsorbed onto the JNS surface as $\kappa$ decreases, see Fig. \ref{fig-profile}A.

As illustrated in Fig.~\ref{fig-fe}, there occur pronounced changes of the binding
energy when transitions between the PE adsorbed and desorbed states in the vicinity
of the JNS occur. More specifically, the inset of Fig.~\ref{fig-fe} shows the time
trace of the PE-JNS binding energy $E_B(t)$ in a given simulations run, while the
main panel of Fig.~\ref{fig-fe} illustrates the long-time equilibrium probability
distribution $p(E_B)$ of the adsorption energy $E_B$. We observe that PE chains at
progressively higher salt concentrations have, as expected, smaller magnitudes
$|E_B|$ of ES attraction to the JNS, see the differences between the circles,
squares, and triangles in Fig.~\ref{fig-fe}, for increasing salinity. At low salt each PE monomer in the bound state has a binding energy of about one unit of
the thermal energy $k_BT$, which corresponds to the ratio of the $E_B$ values of
Fig.~\ref{fig-fe} versus the number of monomers, $N=20$. The distributions $p(E_B)$
are normalised to unity when the contributions of both adsorbed and desorbed PE
states, with zero binding energy, are accounted for. The latter are represented
by the vertical lines at $E_B=0$ in Fig.~\ref{fig-fe}. As expected, at lower salt
and thus stronger ES attraction, the proportion of bound PE states increases, as
given by the relative area under the distributions $p(E_B<0)$ in Fig.~\ref{fig-fe}.

In Fig.~\ref{fig-profile} we show the non-normalised radial distribution function
of the chain monomers facing the JNS (panel A) and a uniformly charged sphere with
the same $\sigma$ (panel B). We obtain a systematically broader distribution of
monomers in front of the JNS compared to the sphere, consistent with the behaviour
of the surface layer width, as shown in Fig.~\ref{fig-thickness}. With addition of salt, the profiles of PE chains adsorbed onto JNSs become progressively broader, as intuitively
expected, and also more monomers are getting adsorbed, as illustrated in Fig. \ref{fig-profile}A. 

Note that the PE monomer profiles in Fig.~\ref{fig-profile} are shown for the
parameters well above the critical adsorption point. At the critical point itself, when
$\sigma\to\sigma_c$, the width $w$ of the polymer distribution diverges. For large
surface charge densities, in contrast, the relation $w(\sigma)\sim\sigma^{-\gamma}$
with $\gamma\approx1/3$ is valid, as shown for PE chains adsorbed onto planar and
curved interfaces in Refs. \cite{Cherstvy:2011,Winkler:2013}.

\subsection{Adsorbed layer width $w$}

\begin{figure}
\includegraphics[width=7cm]{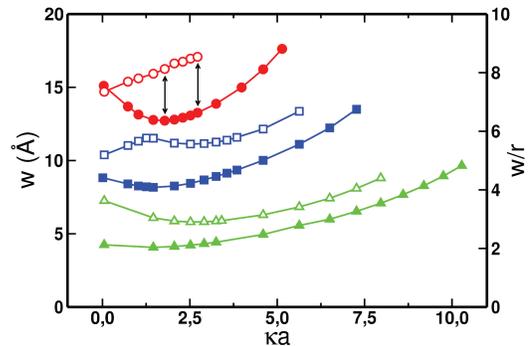}
\caption{Dependence of the width of the layer of the adsorbed PE monomers
at the JNS surface (open symbols) and the surface of a uniformly charged sphere
(filled symbols) versus the salinity of the electrolyte solution. Parameters: $\sigma=12.5$
 (\textcolor{black}{red circles}), 25 (\textcolor{black}{blue squares}), and 50 mC/m$^2$ (\textcolor{black}{green triangles}). We use a
constant JNS radius $a=70$ \AA~ and PEs with $N=100$ monomers. Approximately $10^6$ polymer
conformations were used in the averaging. The arrows show the salinities for the
PE profiles presented in Fig.~\ref{fig-profile}.}
\label{fig-thickness}
\end{figure}

\begin{figure}
\includegraphics[width=6cm]{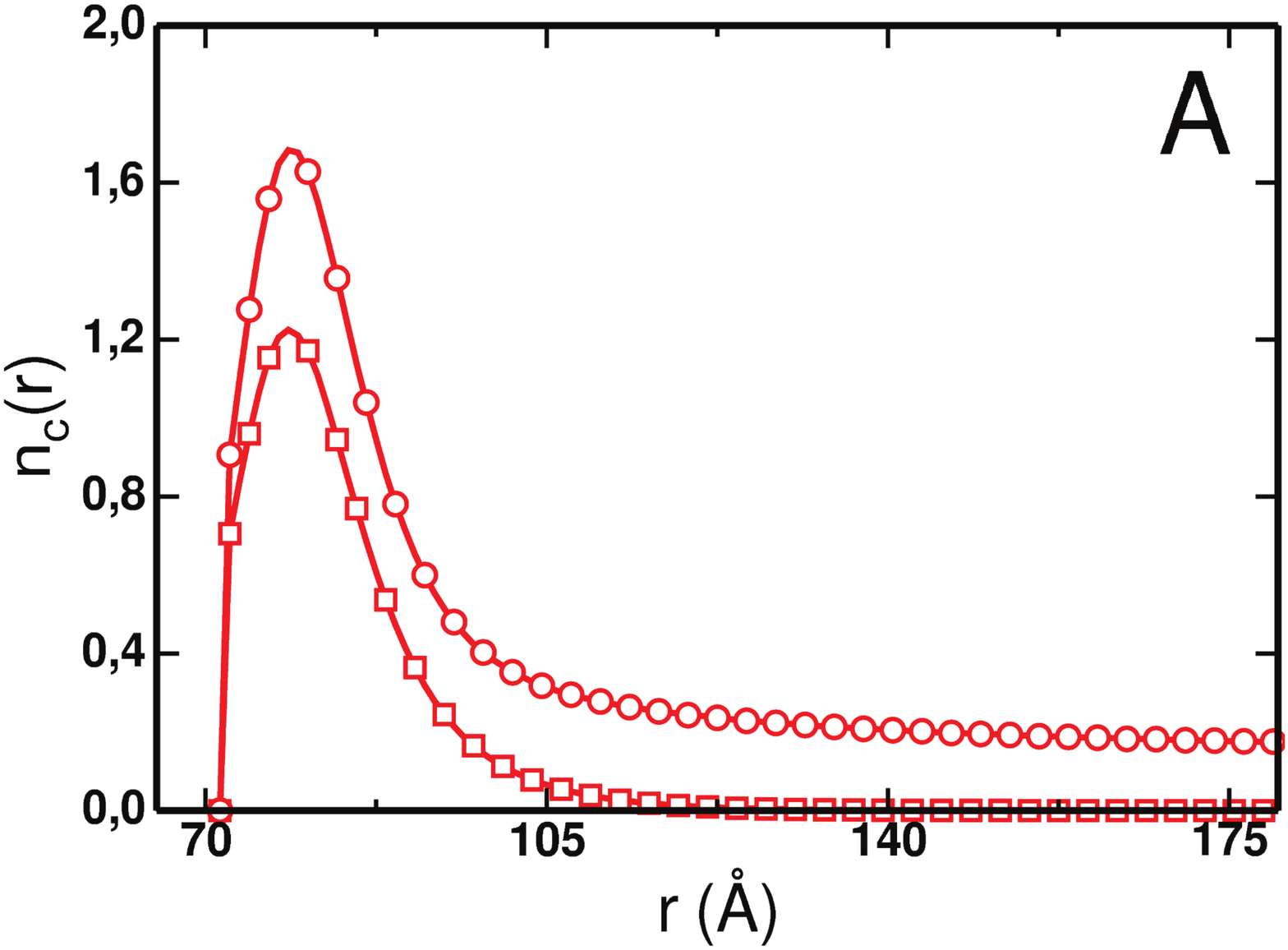}
\includegraphics[width=6cm]{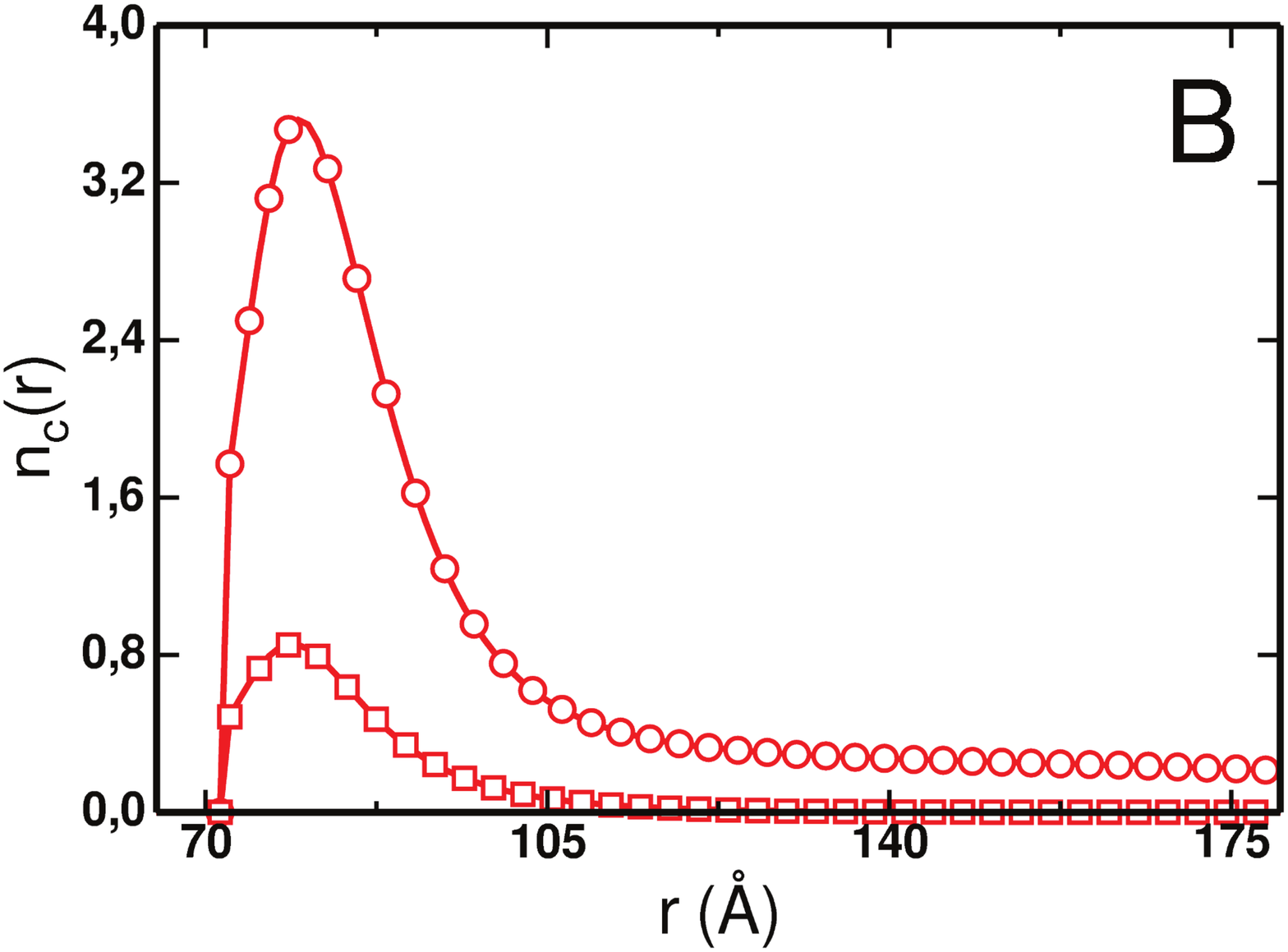}
\caption{Radial distributions of PE monomers in front of JNSs ar  $\sigma=12.5 $ mC/m$^2$ (\textcolor{black}{reddish colors}) and salt concentration of 0 (panel A) and 6 mM (panel B). Other parameters are: $N=100$ (open circles) and $N=20$ polymer monomers (open squares), and JNS radius is $a=$70 \AA.}
\label{fig-profile-length}
\end{figure}

From Fig.~\ref{fig-profile} we obtain the width $w$ of the adsorbed PE layer at
the half-height of the monomer density profile $n_c(r)$. The width typically decreases for
decreasing salinity, as shown in Fig.~\ref{fig-thickness}. This decrease occurs
because of the stronger ES PE-particle attraction, both for PEs in the proximity
of the JNS and the uniformly charged sphere (apart from a region of small values
of $\kappa a$). For the same physical reason, the PE layer width $w$ decreases
with $\sigma$, compare the curves with the open symbols in Fig.~\ref{fig-thickness}.

The highest $\kappa$ values in the curves of Fig.~\ref{fig-thickness} correspond
to the largest salinity of the solution at which PE-particle adsorption
occurs (critical point). At even larger salt concentrations we have two distinct monomer populations: one of them is for PE chains staying near the particle corresponding to the adsorbed state, while another one is for PE\ monomers away from the Janus particle (the desorbed state). The range of salinities in terms of $\kappa a$ in which the PE-JNS adsorption
occurs is systematically \textit{smaller\/} than that for the uniformly charged
sphere, compare the open and filled symbols in Fig.~\ref{fig-thickness}. As
expected, the PEs in front of the JNS give rise to somewhat \textit{thicker} adsorption layers
as compared to the PE adsorption onto the homogeneous sphere at the same conditions,
as shown in Fig.~\ref{fig-thickness}. Another expected observation is the fact
that for smaller values of Janus particle charge density $\sigma$, the region of $(\kappa a)$ values of PE-JNS
adsorption shrinks and shifts towards smaller $\kappa a$ values. Therefore, for
weakly charged JNSs even moderate salt conditions can prevent PE
adsorption. 

We also find that for shorter PEs experiencing weaker attraction to JNSs, the width $w$ of the adsorbed PE layer for the surface charge densities
given in Fig.~\ref{fig-thickness} decreases, cf. Fig. \ref{fig-profile-length}. Physically, longer polymers do experience a stronger ES attraction and their monomers adsorb onto the JNSs in larger numbers. Because of a limited nanoparticle surface available to PE deposition, however, longer PE chains crowd near the attractive JNS hemisphere. The chains therefore swell because of inter-monomer ES\ repulsion. All this  gives rise to larger thicknesses of the adsorbed PE layer as $N$ grows, see Fig. \ref{fig-profile-length} for two values of the solution salinities.

We note that, unlike the planar and cylindrical cases for which the available
surface for PE adsorption is infinite, for spheres and JNSs it is strictly
bounded. This effects the extensive polymer \textit{tail formation}, seen also in the sample configurations in Fig.~\ref{fig-config}. For this reason, the number of monomers in the adsorbed layer at the JNS surface
changes with the ionic strength, resulting in the complex behaviour presented
in Fig.~\ref{fig-thickness}. 

\begin{figure}
\includegraphics[width=7cm]{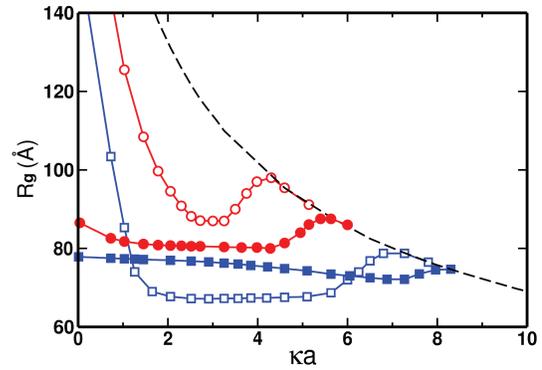}
\caption{Radius of gyration of the PE chain adsorbed onto a uniformly-charged
sphere (filled symbols) and a JNS (open symbols). Parameters, symbol notations,
and colour scheme are the same as in Fig.~\ref{fig-thickness} ($a=70$\AA \ and $\kappa$ is varied). The dashed
black curve indicates the size of a free semiflexible polymer with the proper $l_p$ in solution.}
\label{fig-Rg}
\end{figure}

Finally, in particular for strongly charged JNS,
corresponding to the green triangles in Fig.~\ref{fig-thickness}, we observe a
non-monotonic dependence of the adsorbed layer width on the salt concentration.
This is indicative of a weaker ES attraction of PE chains to the JNS at low salt
conditions, when the effects of the similarly charged JNS hemisphere is amplified,
compensating the PE attraction to the similarly-charged JNS hemisphere. We refer
also to Ref.~\cite{Forsman:2012} which investigates non-monotonic effects of added
salt and non-electrostatic PE-surface interactions analytically and by simulations.

The variation of the radius of gyration $R_g$ of the PE chain during the
adsorption onto JNSs and the uniformly charged spheres is shown
in Fig.~\ref{fig-Rg}. The slightly non-monotonic variation observed in this
plot is quite sensitive to the system parameters, such as the particle size, its
surface charge density, and the polymer length. The universal feature we want to
highlight here, however, is the general dependence of $R_g$ on the salt
concentration. In particular, for salinities above the critical adsorption
transition, the chain size follows the prediction for the non-confined semiflexible polymer
in solution, both for the JNS and the uniformly charged sphere, as shown by the
dashed curve in Fig.~\ref{fig-Rg}. For very low salt concentrations, in contrast,
the gyration radius of PE chains in front of the JNS starts to grow dramatically
as the salinities drops below the critical salinity required for PE-JNS
adsorption. This is indicative of stronger ES repulsion of PE chains from the similarly-charged JNS hemisphere.
This repulsion is long-ranged in the low-salt limit and has a tendency
to displace the polymer from the favorable JNS hemisphere as well as
to stretch the chain. This stretching effect is shown on the left panel of
Fig.~\ref{fig-config} for vanishing salinity.

\subsection{Critical PE adsorption conditions} 
\label{sec-crit}

\begin{figure}
\includegraphics[width=7.1cm]{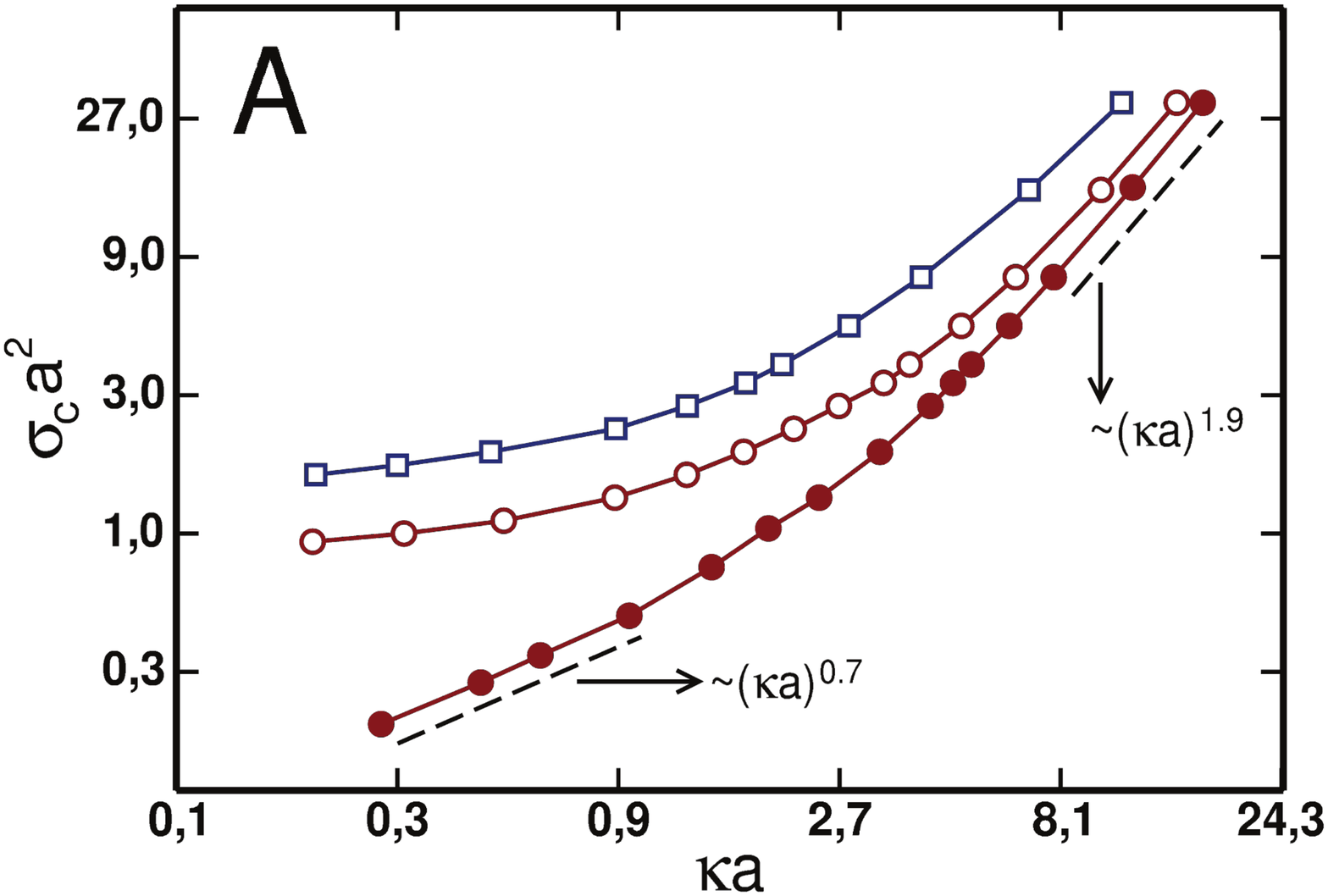}
\includegraphics[width=7.1cm]{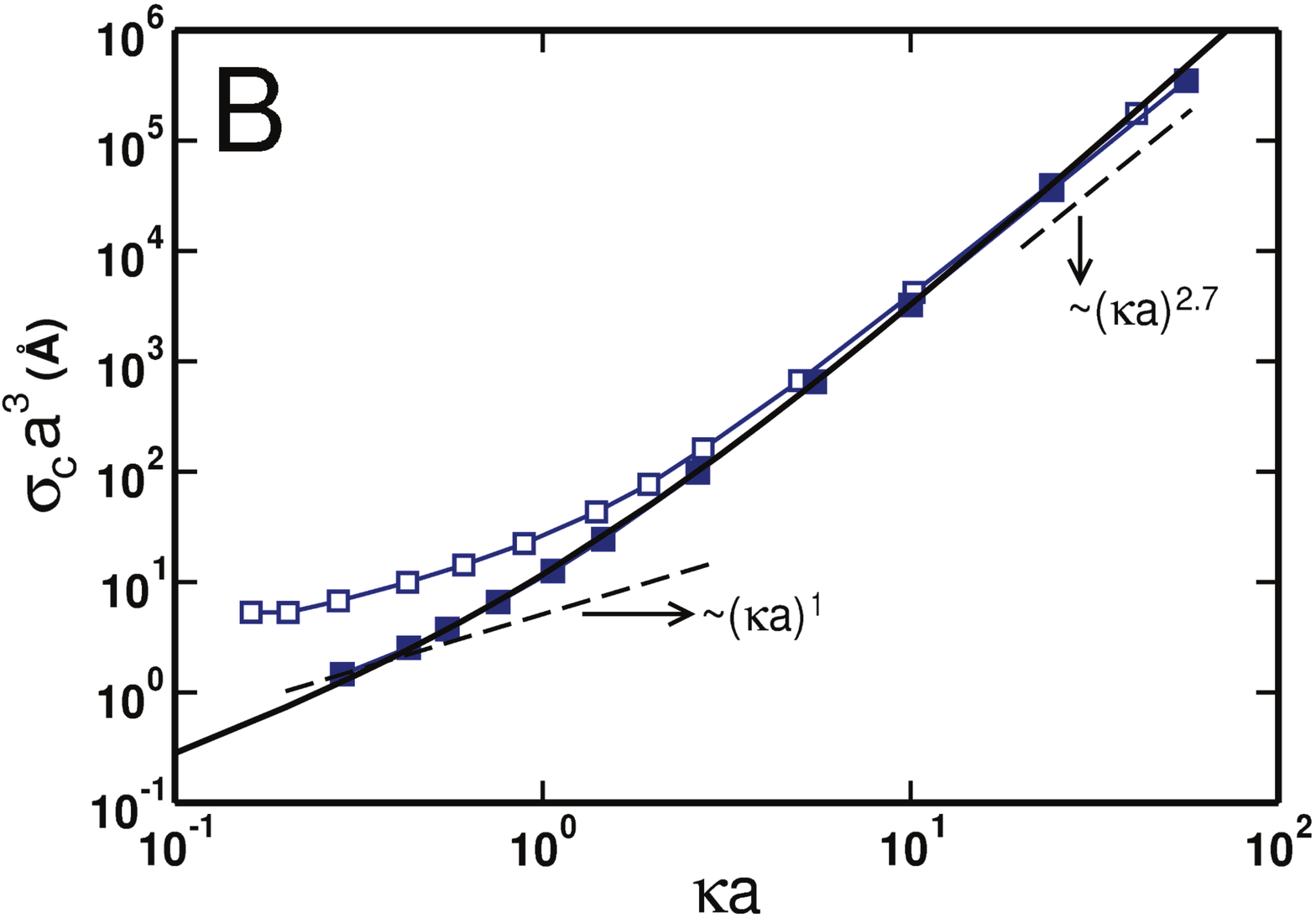}
\caption{A: Variation of salinity. Results of computer simulations for the
critical surface charge density $\sigma_c$ of a JNS-particle  for varying $\kappa$.
Parameters: the constant JNS radius $a=70$ \AA, $N=20$ chain monomers (open squares),
$N=100$ (open circles). The results of the same simulations method for a
uniformly charged sphere are shown as filled circles (polymer length is $N=100$).
B: Variation of the Janus particle radius. The adsorption-desorption threshold at
$\approx$50 mM of 1:1 salt ($\lambda_D\approx 13.6$ \AA) and polymer length $N=20$ (open squares) at varying JNS radius $a$. The results for PE chains with $N=20$ monomers in the proximity of a uniformly charged sphere are represented by the filled squares. The theoretical asymptote of Ref.~\cite{Cherstvy:2011} for PE-sphere adsorption given by Eq.~(\ref{eq-delta_sphere}) multiplied by 1.77 is shown as the solid black curve. On a standard 3-3.5 GHz workstation every point on the graph requires some 17 h and 184 h of computational time for the chains of $N=$20 and $N=$100 monomers, respectively.}
\label{Critical_Cond}
\end{figure}

With our above adsorption criterion, we determine the critical adsorption
conditions for PE chains of different lengths and JNSs of varying sizes. Our
main conclusion is that there exists \textit{no\/} universal scaling of the
parameter $\kappa a$ for PE adsorption onto JNSs, in strong contrast to the
case for PE adsorption onto spherical and cylindrical interfaces
\cite{Cherstvy:2011}. The curvature of the JNS surface and the screening parameter $\kappa$ cannot now be varied inter-changeably. The solution salinity and the surface
curvature affect the dependence of the critical adsorption conditions in a disparate way. This is likely due to the salt effects on the PE persistence
length $l_p$, see Fig.~\ref{Critical_Cond}. Note that the range of $\kappa a$
we present in Fig.~\ref{Critical_Cond} spans more than two decades,
that is significantly more than examined recently for the PE-cylinder adsorption
\cite{deCarvalho:2013}.

Let us now describe the \textit{main results} of the current study presented in
Figs.~\ref{Critical_Cond}A and \ref{Critical_Cond}B. We start with the
results for the uniformly charged sphere and consider a varying sphere radius
$a$, the filled squares in Fig.~\ref{Critical_Cond}B. We find the
scaling $\sigma_c\sim(\kappa a)^{2.7}$ in the limit of large $\kappa a$, in
agreement with the limit of large spheres discussed in Ref.~\cite{Cherstvy:2011}
or with the nearly planar result of Ref.~\cite{Wiegel:1977} given by
Eq.~(\ref{delta_plane}). In the limit of $\kappa a \ll1$ we observe that $\sigma_c
\sim(\kappa a)^1$, in agreement with previous results for small spheres derived in
Refs.~\cite{Cherstvy:2011,Winkler:2006,Winkler:2013} and described by
Eq.~(\ref{delta_sphere}). Note that from the results obtained for the critical
surface charge density $\sigma_c$ and the definition (\ref{def_delta}) one can recalculate
the critical minimal radius $a_c$ of the JNS particle necessary for adsorption. 

\subsubsection{Varying Janus particle size}

For JNSs of varying sizes at constant $\kappa$, we detect a similar turnover
of the scaling behaviour $\sigma_c(\kappa a)$. The critical JNS surface charge density
$\sigma_c$ required to trigger the PE adsorption is \textit{larger\/} than that
for the uniformly charged sphere in the whole range of $\kappa a$, see the empty
and filled symbols in Fig.~\ref{Critical_Cond}B. The $(\kappa a)$ region in
which the change in the scaling behaviour for JNSs from $\sigma_c\sim(\kappa
a)^{1}$ to $\sigma_c\sim(\kappa a)^{\nu}$ with $\nu\approx3$ occurs, is
somewhat \emph{shifted\/} to larger values for JNSs, in comparison to the sphere. For
PE-sphere adsorption, the change in scaling occurs at $(\kappa a)\sim1$, i.e.,
when the Debye screening length is comparable to the sphere size $a$, in agreement with the theory (see Fig.~1 in Ref.~\cite{Cherstvy:2011}) and the experimental
data shown in Fig.~6 in Ref.~\cite{Feng:2001HAHAHA} as well as Fig.~11 in
Ref.~\cite{Cherstvy:2006}.

The theoretically predicted dependence $\delta_c(\kappa a)$ for the PE-sphere
adsorption \cite{Cherstvy:2011} given by Eq.~(\ref{eq-delta_sphere}) and
multiplied by a factor of 1.77 is shown as the solid
black curve in Fig.~\ref{Critical_Cond}B. This numerical factor of the order of unity apparently accounts for all the discrepancies of current simulations performed for finite electrostatically-persistent PE chains versus the infinitely long unperturbed Gaussian polymers considered in the theory. The agreement with the simulations data
both in terms of the magnitude of the critical surface charge density and in terms
of the region in which the transition from the cubic to linear scaling in $\sigma
_c(\kappa a)$ occurs, is remarkably good: the theory agrees with the simulations
over the range of more than two decades of the sphere size. It is important to note that $\delta_c$ contains the effects of a finite PE
persistence and other chain length effects. For this fact the agreement of the
theory \cite{Cherstvy:2011} with our current simulational results for relatively
short chains ($N=20$ for the solid symbols in Fig.~\ref{Critical_Cond}B) is
even more remarkable. This observation is a further support for the quantitative
validity of the simulations results for PE-JNS adsorption while no theory for
this case has been developed to date.

To calculate the error bars for $\sigma_c$, we performed a dozen of simulations with different random number seed to generate different sets of chain configurations. For the most data points in Fig. \ref{Critical_Cond} for $N=20$ chains the deviation in the critical adsorption conditions is just $\sim0.1\dots 1 \%$ of the $\sigma_c$ value. The error bars are thus much smaller than the symbol size. For longer $N=100$ chains the deviations are expectedly larger, namely $\sim5\dots 10 \%$ of $\sigma_c$. The error bars are of the order of symbol size, except for the first couple of points at the smallest solution salinities, when the error bars can be twice the symbol size. This is physically understandable because in the low-salinity regime, close to the complete desorption threshold, the fluctuations of the polymer dimensions are particularly large. For the others quantities we presented, such as the radius of gyration of the chain $R_g$ or the width of the adsorbed PE layer $w$, the error bars are smaller then symbol size.

\subsubsection{Varying salt concentration}

We now turn to the description of the results obtained when the salinity of the
solution is changed in the simulations, as shown in Fig.~\ref{Critical_Cond}A. For
a constant radius $a$, the PE chain in front of the uniformly charged
sphere exhibits the scaling $\sigma_c\sim(\kappa a)^{1.9}$ in the limit of
$\kappa a \gg 1$ and the scaling $\sigma_c\sim(\kappa a)^{0.7}$ for $\kappa a
\ll 1$, in a drastic contrast to Fig. \ref{Critical_Cond}B where the particle
size was varied. The scaling exponent $\nu\approx1.9$ in the limit of $\kappa a\gg 1$,
which is smaller by about unity than that in the setup with constant salt
concentration, is consistent with the effect of the ES PE persistence
length $l_p^{ES}$ on the critical adsorption conditions. This effect is
neglected in the simplistic adsorption theory of flexible Gaussian chains
\cite{Winkler:2013} for which the scaling $\sigma_c\sim(\kappa a)^3$ is effected,
as prescribed by Eq.~(\ref{delta_plane}). For flexible PEs the scaling
\begin{equation}
l_p^{ES}(\kappa)\sim \kappa^{\mu} \mbox{ with } \mu\approx-1
\end{equation}
is known to hold. According to Eq.~(\ref{def_delta}) and with the total PE
persistence length $l_p=b/2=l_p^{0}+l_p^{ES},$ the renormalised $\sigma_c\sim(
\kappa a)^2$ scaling for $\kappa a\gg1$ is obtained for the critical adsorption
of PEs with such a $\kappa$-dependent persistence length onto a spherical particle, as we indeed
observe in the simulations when  $\kappa$ is varied. 

In the limit $\kappa a\ll1$ the scaling exponent in the setup with a varying salt
concentration is also somewhat reduced for the sphere, namely down to  $\nu\approx$0.7, see the filled symbols in Fig.~\ref{Critical_Cond}A. For small
$\kappa a$, overall much smaller values of $\sigma$ are required to trigger the
PE adsorption onto a sphere, as compared to JNSs. This is due to the stronger
PE-sphere attraction which overcomes the entropic penalty of the polymer
confinement near the adsorbing surface. For the PE-adsorption onto a JNS at
varying $\kappa$ the overall growth of $\sigma_c$ with $\kappa a$ at $\kappa a\ll
1$ is much weaker than for the sphere, as demonstrated by the open symbols in
Fig.~\ref{Critical_Cond}A. For $\kappa a\ll1$ the critical surface charge density
for PE-JNSs adsorption almost saturates, that is, $\sigma_c\sim\kappa^\nu$ with $\nu \ll\ 1$. With the adsorption criterion used here we do not observe any increase of the
value of $\sigma_c$ at very small $\kappa a$, but we cannot exclude this for
modified/different adsorption conventions. 

\begin{figure}
\includegraphics[width=7cm]{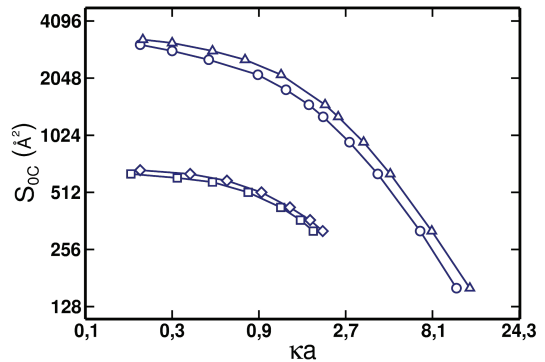}
\caption{Critical area $S_{0}=e_{0}/|\sigma_c|$ per surface charge on a Janus particle for PE-JNS adsorption as function of the surface curvature and the salinity. The results are obtained with the exact ES potential $\Psi_J(r,\theta)$ (circles
and squares) and the $l=0,1$ terms in Eq.~(\ref{exact_potential}) (triangles
and rhombi symbols). The solution salinity is varied while the other parameters
are $N=20$, $a=70$ \AA ~(upper curves) and $a=30$ \AA ~(lower curves).}
\label{Approx}
\end{figure}

Also note that in the full nonlinear ES
model of Eq.~(\ref{eq-nlpb}) \emph{lower\/} ES potentials will occur
near the charged interfaces for the same $\sigma$, as compared to the linear
ES theory of Eq.~(\ref{eq-lpb}). Then, to ensure the same PE-surface affinity
(\ref{eq-ebinding}), effectively \emph{higher\/} surface charge densities
will be necessary. This will serve as another source for a potential \textit{increase}
of $\sigma_c$ for realistic Janus particles at very low salt conditions,
where the nonlinear ES model has to be applied.      

Moreover, with the chosen adsorption criterion, longer PE chains which are
attracted stronger to the particle surface need weaker Janus
hemisphere charge densities $\sigma_c$ for the  adsorption to occur, compare
the open squares ($N=20$) and circles ($N=100$) in Fig.~\ref{Critical_Cond}A.
This trend is similar to that for PE-sphere adsorption for varying polymerisation
degree of the PE chain, compare Fig.~5 of Ref.~\cite{deCarvalho:2010}.

Finally, note that for the critical adsorption conditions presented in Fig. \ref{Critical_Cond} we implemented the
full exact ES potential (\ref{exact_potential}). To save simulation time, it is also possible to use just the first two terms in the ES potential expansion
(\ref{exact_potential}). The obtained results are quite close to the exact
ones, over a broad range of parameters, see Fig.~\ref{Approx}. In this figure
we present the critical surface area $S_0$ per charge that is a directly
measurable experimental parameter: $S_{0,c}=e_0/|\sigma_c|$.

\section{Discussion and Conclusions}
\label{Sec_Disc}

Let us first summarise our main findings. From extensive computer simulations we
uncovered the properties of adsorption of flexible PE chains onto net-neutral
Janus nanospheres with oppositely charged hemispheres. The adsorbing PE chain
occupies the JNS hemisphere of the opposite charge and avoids the like-charged
hemisphere of the JNS. We examined the behaviour of the critical adsorption
conditions in dependence of the particle surface charge density, the surface
curvature, and the length of the PE chain. The latter dependence is particularly
hard to predict theoretically. We demonstrated that the PE-JNS
adsorption-desorption transition differs in many respects from PE
adsorption onto the uniformly charged sphere. 

In particular, we discovered that there exists \textit{no\/} universal parameter
$\kappa a$ that couples the surface curvature and the salinity of the solution.
The results for the critical PE-JNS adsorption conditions differ substantially
if one separately varies the salt concentration or the JNS radius, as shown in
Fig.~\ref{Critical_Cond}. We rationalised the differences in these scalings using
the concept of a salt-dependent PE persistence. We also found that shorter
PE chains necessitate larger charge densities on the JNS hemispheres to trigger
adsorption. We explored how the width of the adsorbed PE layer is
reduced for JNSs with progressively larger surface charge densities,
due to stronger ES adsorption. For JNSs the critical surface charge density and
the thickness of adsorbed layer throughout remain larger than for the PE
adsorption onto uniformly charged spherical particles of the same size and
nominal surface charge. This reflects the influence of the JNS hemisphere
of the same  charge as the PE chain, which opposes the adsorption via
repulsive ES JNS-PE interactions.      

A direct biological application of our results for PE-JNS interaction is the
description of the critical adsorption conditions of flexible and semi-flexible
PE chains onto oppositely charged interfaces of spherical and cylindrical geometry
that have been thoroughly studied in the lab of Paul Dubin \cite{Cooper:2005,Chen:2011,
Feng:2001HAHAHA,Dubin:2001powers,Kizilay:2011,Kayitmazer:2013,Dubin:1992}.
Flexible PE chains in the experiments (such as PAMPS, polyacrylic acid PAA, random
copolymers of 2-(acrylamido)-2-methylpropane-sulfonate and acrylamide P(AMPS-AAm),
sulfonated poly(vinyl alcohol) PVAS, sodium-poly(styrene sulfonate)
NaPSS, or poly(dimethyldiallylammonium chloride) PDADMAC) usually have
persistence lengths of the order of $l_p\approx 1\dots4$ nm, whereas for semiflexible
chains such as double stranded DNA the persistence length is of the order of  $l_p\sim50$ nm.
The oppositely charged surfaces often include dodecyldimethylamine oxide DMDAO
cationic uniformly charged micelles with well titration-controlled surface charge
density, controllable dimensions, and a geometry which depends on the salt
conditions (spherical micelles occur at low salt and cylindrical ones at high
salt conditions). Other complexes include spherical colloidal particles,
dendrimers of varying generations, and globular proteins \cite{Kayitmazer:2013}
that are often non-homogeneously charged (e.g., BSA bovin serum albumin and lysozyme).

In brief, experimentally the complexation of PEs with oppositely charged particles
is measured via the changes in the solution turbidity. The latter is intimately
coupled to coacervation, aggregation, and precipitation of complexes. The onset on
complex formation is well controlled and precisely measured. Turbidimetric
measurements systematically reveal that with increasing solution salinity the
critical surface charge density $\sigma_c$ of the particle surface grows, as seen from
Eq.~(\ref{eq-dubin-scaling}). The complex formation of flexible PVAS polymers
with nearly uniformly charged DMDAO micelles yields scaling exponents of $\nu
\approx$1.0 and $\nu\approx$1.8 for spherical and cylindrical cases, respectively
\cite{Feng:2001HAHAHA}. For P(AMPS-AAm) complexes with DMDAO micelles the powers
are $\nu\approx$1.4 and $\nu\approx$2.5 \cite{Feng:2001HAHAHA}. DNA complexation
with DMDAO micelles of spherical and cylindrical geometry reveals the exponents of
$\nu\approx$1.6 and $\nu\approx$1.8, respectively \cite{Dubin:2001powers}.

\begin{figure}
\centering
\includegraphics[width=8cm]{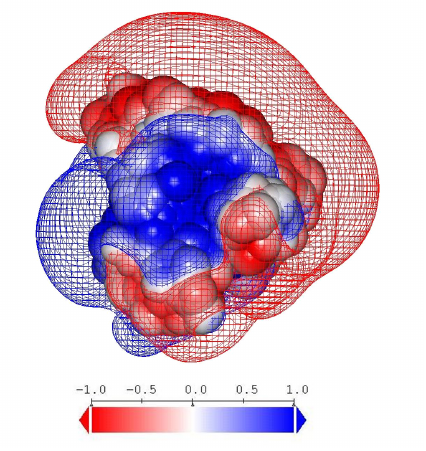}
\caption{ES potential of the BSA protein, as simulated in Ref.~\cite{Cooper:2006}
at pH=6.2 and ionic strength of 0.05 M. The blue colour corresponds to a positive
potential (a cationic patch). The mesh illustrates the surfaces with the dimensionless
potential of $|e_0\Psi/(k_BT)|=$0.1. The red-blue scale denotes the magnitude $|e_0
\Psi/(k_BT)|$ of the potential. The image is from Fig.~11A of
Ref.~\cite{Cooper:2006}, subject to ACS-2006 Copyright. Original figure file
courtesy S. Stoll.}
\label{fig-bsa}
\end{figure}

Charge anisotropies of heterogeneously charged proteins such as BSA can even
further reduce the scaling dependence of the critical surface charge density
for adsorption with $\kappa$. Note that for patchy proteins, to compute the effective surface charge density for adsorption $\sigma_c$, the net protein charge was divided over the protein surface. Indeed, the complexation of a number of poly-anions (AMPS/AAm, PAA, pectin)
with spherical cationic micelles, dodecyltrimethylammonium bromide/Triton X-100
DTAB/TX-100, revealed the following critical exponents for the critical surface
charge density $\sigma_c\sim\kappa^\nu$: for uniformly charged micelles $\nu
\approx1.7$ versus $\nu\approx0.9$ for pronouncedly heterogeneous charge
distribution of BSA proteins. The BSA charge distribution, with a single
relatively large cationic patch present at low salt, was additionally shown to evolve
into multi-patchy structure at higher salinities, see Fig.~\ref{fig-bsa} and
the original Fig.~11 in Ref.~\cite{Cooper:2006}. 

Thus, at low salinities the anionic part of the BSA surface impedes the poly-anion
adsorption onto the protein surface, repelling it electrostatically from the
attractive cationic patch and giving rise to dramatically smaller values of the
critical exponent $\nu$. These experimental observations therefore very nicely
support our predictions for the PE conformations in the proximity of the JNSs
in Fig.~\ref{fig-config} and the much weaker growth of $\sigma_c$ with $\kappa$
for PE adsorption onto JNSs, see the results of Fig. \ref{Critical_Cond}. The JNSs are somewhat reminiscent in their charge surface pattern to BSA at low
salt conditions, as illustrated in Fig.~\ref{fig-bsa}. We also note that a number
of PE complexes with patchy proteins revealed a maximal affinity as function of
the solution salinity realised for Debye screening lengths comparable to the
protein dimensions \cite{Cooper:2006}. 

Another example of PE-protein complexation is the ES-driven wrapping of a
negatively charged double-stranded semi-flexible DNA fragment around highly basic histone-core
proteins in nucleosome core particles \cite{Schiessel:2003}. The latter exhibit an equatorial ring of
positive Lysine and Arginine charges \cite{Luger:1997,Cherstvy:2009} forming
the strongly non-uniformly charged substrate for the binding of DNA. In both
cases, for PE-BSA and DNA-histone adsorption, the surface charge patchiness and structure can give rise to PE adsorption
on the \textit{wrong\/} side of the isoelectric point and to complex overcharging
\cite{Netz:2014} (see, however, also Ref.~\cite{Yoshi:2005}). One more example of
patterned surfaces is the adsorption of DNA onto laterally-structured interfaces
of supported cationic lipid membranes, as monitored by atomic force microscopy
techniques in Ref.~\cite{Clausen-Schaumann:1999}, see also Refs.~\cite{Farago:2006}
and \cite{Petrov:2014} for, respectively, computer-based and theoretical
modelling of DNA complexation with cationic lipid membranes.

In the case of flexible PE-surface adsorption, the ES-driven ordering of flexible
single-stranded RNA molecules on highly positively charged interiors of viral
capsids was suggested \cite{Muthu:2006,Muthu:2009,Rudi:2012viruses}. The RNA
adsorption in this system takes place onto strongly non-uniformly charged,
patterned surfaces decorated with flexible, highly-basic polypeptide arms. The
latter grab the RNAs and trigger mutual complexation, a prerequisite for
sustainable thermodynamic self-assembly of single-stranded RNA viruses from the nucleic acid and
the protein-subunit components available in the solution \cite{Hagan:2013}. The
resulting structure pattern of single-stranded RNA molecules adsorbed to the inner
capsid surface obtained both from computer simulations and from experiments
indeed resembles the symmetry of viral shells \cite{Muthu:2009}. 

Extensions of the current simulations scheme will include PE adsorption onto
charge-corrugated and patterned Janus surfaces of spherical and cylindrical
shape. Also, semiflexible PEs and weak versus strong adsorption conditions will
be studied. The latter will e.g. lead to the formation of some ordered, snail-like or
toroidal PE structures on oppositely charged caps of the Janus particles, in
contrast to the quite disordered PE appearance of flexible chains weakly
adsorbed on JNSs, compare Fig.~\ref{fig-config}. Ordered PE structures, such as
tennis-ball like, rosettes, and solenoids, are known to appear in the ground state
upon strong PE adsorption onto oppositely and uniformly charged spherical particles, see, e.g.,
Refs.~\cite{Cherstvy:2006,Schiessel:2003,Schiessel:2000,Netz:2005,Netz:1999}.
Lastly, the effects of PE charge heterogeneity and patchiness should be considered
in the future, as it is known to affect the critical adsorption conditions as well. 

It will be important to test the theoretical results presented here by experimental
studies of the different adsorption characteristics of PE chains onto Janus-like
nanoparticles, as compared to PE adsorption onto uniformly charged convex surfaces.

\section{acknowledgments}
We thank Paul Dubin for insightful comments regarding the PE binding to BSA
proteins and to Serge Stoll for kindly providing the original images of the BSA
ES surface potential. AGC acknowledges a fruitful collaboration with Roland
Winkler over many years. The authors acknowledge financial support from the
Academy of Finland (FiDiPro scheme to RM) and the German Research Foundation (DFG
grant CH 707/5-1 to ACG). Computer resources were supplied by the Center
for Scientific Computing (NCC/GridUNESP) of the S{\~a}o Paulo State University.

\appendix

\section{ES potential of the Janus particle}
\label{sec-app-es}

We here derive the ES potential around a JNS. The charge density of a JNS of
radius $a$ with the charge density $\pm\sigma_0=\pm e_0/S_0$ on its caps in
spherical coordinates can be written as
\begin{equation}
\sigma(r,\theta)=\sigma_0\delta(r-a)\Big[\Theta(\pi/2-\theta)-\Theta(\theta-
\pi/2)\Big].
\end{equation}
Here $\theta$ is the polar angle, $\Theta(x)$ is the step function, $e_0>0$ is
the elementary charge, and $S_0$ is the area per charge on the surface. The
linear Poisson-Boltzmann equation for the ES potential $\Psi_{J}$ of a JNS in
an electrolyte solution,
\begin{equation}        
\label{eq-lpb}
\nabla_{r,\theta}^2 ~\Psi_{J}(r,\theta)=\kappa^2\Psi_J(r,\theta),
\end{equation}
allows the separation of variables via the eigenfunction expansion in terms of
Legendre polynomials $P_l(x)$ and spherical Bessel functions $K_{l+1/2}(x)$
\cite{Cherstvy:2006}, namely  
\begin{equation}
\Psi_{J}(r,\theta)=\sum_{l=0}^{\infty}C_l P_l(\cos\theta)\frac{K_{l+1/2}(\kappa r)}
{\sqrt{\kappa r}}.
\label{exact_potential}
\end{equation} 

Using Gauss' theorem on the particle surface, $\partial\Psi_{J}(r,\theta)/\partial
r|_{r=a}=-4\pi\sigma(a,\theta)/\epsilon$, one can restore the distribution of the
ES potential, yielding the coefficients of the eigenfunction expansion
\begin{equation}
C_l=\frac{2\pi(2l+1)}{\epsilon k_l(a)}\int_0^\pi P_l(\cos\theta)\sigma(a,\theta)
d(\cos\theta), 
\end{equation}
with $k_l(a)=\left[2\kappa aK_{l+1/2}'(\kappa a)-K_{l+1/2}(\kappa a)\right]/
\left[2a\sqrt{\kappa a}\right]$ and $K'(x)$ denoting the derivative. Naturally,
the potential $\Psi_{sp}(r)$ of a uniformly charged sphere is given by the $l=0$
term of this expansion, obtained for $\sigma(r,\theta)=\sigma_0\delta(r-a)$ as
\begin{equation}
\Psi_{sp}(r)=4\pi\sigma_0a^2\frac{\exp(-\kappa[r-a])}{\epsilon r(1+\kappa a)}.
\label{eq-pot-sphere}
\end{equation}
After integration of $P_l(x)$, for a JNS we obtain
\begin{equation}
C_l=-\frac{\sigma_0(2l+1)\left(1-\left(-1\right)^l\right){\pi}^{3/2}}{\epsilon
k_l(a)\Gamma(1-l/2)\Gamma(3/2+l/2)},
\end{equation}
where $\Gamma(x)$ is the Gamma function. This yields the ES potential near
JNSs used in the simulations. A contour plot of the potential is shown in
Fig.~\ref{dipole}.
 
\begin{figure}
\includegraphics[width=7cm]{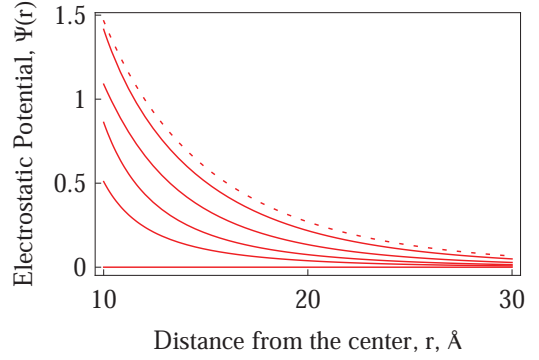}
\caption{Dimensionless ES potential of a JNS, for polar angles $\theta=0$, $\pi
/3$, $5\pi/12$, $11\pi/24$, and $\pi/2$, for the red solid curves from top to
bottom. The dotted curve is the sphere potential $\Psi_{sp}(r)$ given by
Eq.~(\ref{eq-pot-sphere}), for identical radius and $\sigma$. Parameters are
the same as in Fig.~\ref{dipole}.}
\label{potential}
\end{figure}
  
As expected, for JNSs the ES potential shown in Fig.~\ref{potential} is always
lower in magnitude than for a sphere with uniform charge distribution but the
same $\sigma$. For large JNSs, taking along sufficiently many terms in
Eq.~(\ref{exact_potential}), one arrives at the intuitive result that at zero polar angle the asymptotic relation
\begin{equation}
\Psi_{J}(r,\theta=0)\to\Psi_{sp}(r)
\end{equation}
hold. The dependence on $\theta$ illustrated in Fig.~\ref{potential} shows that
close to the ``poles'' of the JNS the deviation from $\Psi_{sp}$ is relatively
small, while close to the ``equator'', i.e., when $\theta\to\pi/2$, the ES
potential is strongly diminished by mutual compensation of the ES effect of the two
oppositely-charged hemispheres. 

Note here that when we vary the salinity of the solution, it is important to
remember that at low salt concentrations the linear Poisson-Boltzmann theory
used for the derivation of the potential in Eq.~(\ref{exact_potential}) breaks
down due to the high ES potentials near the interface. Therefore, the entire
linear ES approach becomes inapplicable in this range. Numerical computations
of the ES potential emerging from JNSs in the full non-linear model,
\begin{equation}
\nabla_{r,\theta}^2\Psi_{J}(r,\theta)=\kappa^2\sinh[\Psi_J(r,\theta)],
\label{eq-nlpb}
\end{equation}
therefore need to be performed. This is a target for future investigations of
more realistic PE-JNS adsorption. We also plan to examine more systematically
the effects of the length as well as the linear charge density and the charge
patchiness of the PE chain.

\section{2D WKB scheme for PE adsorption}
\label{Sec_WKB}

We here present the basic equations for the theoretical analysis of the PE-JNS
adsorption. Using the WKB method for uniformly charged interfaces established in
Ref.~\cite{Cherstvy:2011}, one can, in principle, investigate the adsorption of
a long PE chain onto a JNS. The equation for the polymer eigenfunctions $\psi(r,
\theta)$ in the external Debye-H{\"u}ckel potential field $V_{DH}(r)$ can be
derived from the Edwards equation \cite{Winkler:2006}
\begin{equation}
\left(\frac{\partial}{\partial L}-\frac{b}{6}\nabla_{\mathbf r}^2+\frac{V_{DH}
(\mathbf r)}{k_B T}\right)G(\mathbf{r},\mathbf{r'}; L)=\delta(\mathbf{r}-\mathbf{
r'})\delta( L), 
\label{eq-edw}
\end{equation}
written for the polymer's Green function, expanded in terms of eigenfunctions
\begin{equation}
G(\mathbf{r},\mathbf{r'}; L)=\sum_n\psi_n^*(\mathbf{r'})\psi_n(\mathbf{r})
e^{\mu_n L}.
\end{equation}
For polymer chains much longer than the Kuhn length, $L\gg b$, the ground state
eigenfunction $\psi_0(e,\theta)$ dominates this expression. With the separation
of variables $\psi_0(r,\theta)=\sum_j{\widetilde{R_j}(r)}r^{-1}T_j(\theta)$,
because of the linearity of Eq.~(\ref{eq-edw}) for each $j$ one finds
\begin{eqnarray}
\nonumber
-\mu T_j(\theta)\frac{\widetilde{R_j}(r)}{r}&=&-\frac{|\rho\Psi_{J}(r,\theta)|}{
k_BT}T_j(\theta)\frac{\widetilde{R_j}(r)}{r}\\
\nonumber
&&\hspace*{-2cm}
-\frac{b}{6}\left\{T_j(\theta)\frac{\partial^2\widetilde{R_j}(r)}{r\partial r^2}
\right.\\
&&\hspace*{-1.6cm}+\left.\frac{\widetilde{R_j}(r)}{r^3}\left(\frac{\partial^2
T_j(\theta)}{\partial\theta^2}+\cot\theta\frac{\partial T_j(\theta)}{\partial
\theta}\right)\right\}
\label{eq-2D-WKB}
\end{eqnarray}
For a uniformly charged sphere, $T_j(\theta)\equiv0$, and for the radial component
we get 
\begin{equation}
-\frac{b}{6}\frac{\partial^2\widetilde{R_j}(r)}{\partial r^2}\frac{1}{r}-\frac{4
\pi|\rho\sigma|}{\epsilon k_BT}a^2\frac{\widetilde{R_j}(r)}{r}\frac{e^{-\kappa(r
-a)}}{r(1+\kappa a)}=-\mu\frac{\widetilde{R_j}(r)}{r}.
\label{eq-wkb-sphere}
\end{equation}
This equation was solved by the WKB method \cite{Cherstvy:2011}, with the critical
adsorption condition as function of the universal parameter $\kappa a$,
\begin{equation}
\delta_c^{sp}(\kappa a)=\frac{6\kappa a(1+\kappa a)C^2}{2\pi e^{\kappa a}\mathrm{
erfc}^2(\sqrt{\kappa a/2})},
\label{eq-delta_sphere}
\end{equation}
where $\mathrm{erfc}(x)=1-\mathrm{erf}(x)$ is the complementary error function
and $C\approx0.973$. In the limit $\kappa a\ll1$ this leads to
Eq.~(\ref{delta_sphere}) in the main text. 
 
The ES potential $\Psi_{J}(r,\theta)$ of a JNS is a sum of coupled radial and
polar angle terms, the first term reading $C_1\cos\theta{K_{3/2}(\kappa r)}/
\sqrt{\kappa r}$. The variables for polymer eigenfunctions in the $(r,\theta
)$-coupled ES field near the JNS cannot be separated, however. Thus, instead
of finding an approximate solution of Eq.~(\ref{eq-2D-WKB}), in the main text we
study the PE-JNS adsorption directly and explicitly by computer simulations. We
exploit the critical adsorption conditions and the effect of the \textit{PE length},
which cannot be envisaged from the ground-state-based analysis typically performed
for the eigenfunction (\ref{eq-wkb-sphere}), which is only applicable to long polymers.

We note that other standard approximations to PE adsorption problems are
discussed in Sec.~IV of Ref.~\cite{Cherstvy:2012dielectric}. We do not consider
the charge regulation and self-consistent charge normalisation effects that
might affect the PE adsorption properties which was experimentally noticed
\cite{Dubin:1997}. No account is taken of changes in the counterion atmosphere
upon PE-particle adsorption and a possible low-dielectric JNS interior,
see, e.g., Ref.~\cite{Sens:2000} for more on these effects for the standard
PE-plane adsorption. Non-ES interactions are not taken into account, see,
e.g., Ref.~\cite{Forsman:2012} for more details.

\bibliography{refs} 

\end{document}